\documentclass[IOP]{emulateapj}
\usepackage[utf8]{inputenc}
\usepackage{natbib}
\begin{document}
\title{Near-IR Direct Detection of Water Vapor in Tau Bo\"{o}tis b}
\author{Alexandra C. Lockwood\altaffilmark{1}, John A. Johnson\altaffilmark{1,6}, Chad F. Bender\altaffilmark{2,3}, John S. Carr\altaffilmark{4}, Travis Barman\altaffilmark{5}, Alexander J.W. Richert\altaffilmark{2,3} and Geoffrey A. Blake\altaffilmark{1}}
\altaffiltext{1}{Division of Geological and Planetary Sciences, California Institute
of Technology, Pasadena, CA 91125}
\altaffiltext{2}{Dept. of Astronomy \& Astrophysics, Pennsylvania State University, University Park, PA 16802}
\altaffiltext{3}{Center for Exoplanet \& Habitable Worlds, Pennsylvania State University}
\altaffiltext{4}{Naval Research Laboratory, Washington, DC 20375}
\altaffiltext{5}{Lunar and Planetary Laboratory, University of Arizona, Tucson, AZ 85721}
\altaffiltext{6}{Harvard-Smithsonian Center for Astrophysics; Institute for Theory and Computation, Cambridge, MA 02138}
\email{alock@caltech.edu}

\begin{abstract}
We use high dynamic range, high-resolution $L$-band spectroscopy to measure the radial velocity variations of the hot Jupiter in the $\tau$~Bo\"{o}tis planetary system. The detection of an exoplanet by the shift in the stellar spectrum alone provides a measure of the planet's minimum mass, with the true mass degenerate with the unknown orbital inclination. Treating the $\tau$~Boo system as a high flux ratio double-lined spectroscopic binary permits the direct measurement of the planet's true mass as well as its atmospheric properties.  After removing telluric absorption and cross-correlating with a model planetary spectrum dominated by water opacity, we measure a 6-$\sigma$ detection of the planet at K$_{p}$ = 111 $\pm$ 5 km$/$s, with a 1-$\sigma$ upper limit on the spectroscopic flux ratio of 10$^{-4}$.  This radial velocity leads to a planetary orbital inclination of $i$ = 45~$^{+3}_{-4}$~degrees and a mass of M$_{P}$ = 5.90~$^{+0.35}_{-0.20}$~M$_{\rm Jup}$.  We report the first detection of water vapor in the atmosphere of a non-transiting hot Jupiter, $\tau$~Boo\,b.
\end{abstract}
\let\subjectheadings=\keywords

\section{Introduction}

Since the first detection of an exoplanet around a main-sequence star \citep{Mayor95}, astronomers have discovered hundreds of exoplanets using the radial velocity (RV) technique \citep{Wright12}.  This powerful tool for discovering exoplanets only measures the minimum, or ``indicative" mass, ${M_{\rm p} \sin{i}}$, leaving a degeneracy between two interesting properties of the system, one physical and one orbital.  While the indicative mass is useful for statistical studies, uniquely measuring the true mass of an exoplanet not only yields a key physical property, but also furthers our understanding of planetary formation and evolution via measurement of the true mass distribution of exoplanets  \citep{Zucker01, Weiss13}.  
For example, \citet{HoTurner11} demonstrate that knowledge of the true mass distribution is necessary to convert a minimum mass, ${M_{\rm p} \sin{i}}$, into an estimate of a planet's true mass for RV-detected systems.

One class of objects for which the planet mass can be directly determined are those that transit their host star.  Hundreds of transiting planets have been discovered and characterized, and the ongoing \textit{Kepler} mission has found potential exoplanet candidates numbering in the thousands \citep{Borucki11, Batalha13}.  Transit events also provide atmospheric information through transmission spectroscopy and secondary eclipses.  Investigators have measured spectra of some of the larger transiting planets \citep{Charbonneau02}, leading to the discovery of species such as water, methane and carbon monoxide \citep{Knutson12, Berta12, Crossfield13, Baskin13}.  A variety of spectral retrieval methods have been used to verify the results, confirming the hypothesis that O- and C-bearing gases are present in the atmospheres of hot Jupiters \citep{Madhusudhan09, Line12}. 

Recently, a technique previously used to detect low mass ratio, spectroscopic binary stars has been applied to stars known to host exoplanet systems.  The direct RV detection of an exoplanet involves separating the planetary and stellar components spectroscopically.  The high flux ratio between the primary and companion makes these detections difficult, but not impossible, thanks to modern infrared echelle spectrographs.  \citet{Snellen10} detected CO on HD\,209458\,b with a precision of 2 km/s using the R=100,000 Very Large Telescope CRIRES instrument.  This detection provided the spectroscopic orbit of the system and determined the true mass of the planet.  
      
Indeed, this technique can be applied to non-transiting, RV-detected exoplanets to extract the unknown inclination and true mass.  CRIRES was also used to detect CO on $\tau$~Bo\"{o}tis b \citep{Brogi12}, the first ground-based detection of a short-period non-transiting exoplanet atmosphere, a result confirmed shortly thereafter by \citet{Rodler12}.   These studies provide the true mass of the planet and probe the chemical composition of its atmosphere.  A combination of high signal-to-noise, high spectral resolution, and coverage of multiple CO overtone lines
was required to achieve the sensitivity required for these detections.

Despite the agreement between the two groups, the direct detection of exoplanets, especially $\tau$~Boo\,b, has a storied history.  \citet{CollCam99} first reported a detection of reflected light from $\tau$~Boo\,b more than a decade ago and \citet{Wiedemann01} reported a possible detection of CH$_4$ in the planet's atmosphere soon thereafter, using the same orbital solution.  However, the planetary velocity from the earlier results disagrees with more recent findings discussed above.  

Here we report a detection of water vapor in the atmosphere of $\tau$~Boo\,b, using spectroscopic observations centered around 3.3~$\mu$m. The $\tau$~Bo\"{o}tis system is a stellar binary comprised of a F-type `A' component and a M-dwarf `B' component.  The hot Jupiter $\tau$~Boo\,b orbits the larger `A' component with a period of 3.312 days and a minimum (indicative) mass M$\sin{i}$ = 3.87M$_{Jup}$  \citep{Butler97}. Our detection confirms the mass determined by \citet{Brogi12} and serves to further characterize the atmospheric chemistry of this exoplanet.  Data from five epochs reveal orbital motion of $\tau$~Boo\,b that is consistent with that reported by \citet{Brogi12}. The planetary template spectrum used in the cross-correlation is dominated by water vapor opacity, providing strong evidence of water in the atmosphere of a non-transiting hot Jupiter for the first time. 
  
\section{Methods}

\subsection{Observations and Data Reductions}

We observed the $\tau$~Boo system on five separate nights in March 2011, May 2011, and April 2012 (Table \ref{table:tau_boo_obs}), chose to optimize phasing near orbital quadrature, allowing for maximum separation of stellar and planetary lines.  We use the Near Infrared Echelle Spectrograph (NIRSPEC) \citep{Mclean95} at the W.M. Keck Observatory, which provides high resolution (R $\sim$ 25,000 for a 3-pixel slit) in multiple orders at the wavelengths of interest to study near-infrared water emission.  We obtained spectra covering 3.404-3.457 $\mu$m, 3.256-3.307 $\mu$m, 3.121-3.170 $\mu$m, and 2.997-3.043 $\mu$m. Each epoch covers a total elapsed time of approximately one hour, and is comprised of a continuous sequence of hundreds of exposures.

We extract NIRSPEC spectra using a custom Interactive Data Language (IDL) optimal extraction pipeline, similar to that described by \citet{Cushing04}.  The spatial profile weighting intrinsic to optimal extraction provides a reliable method for detecting and removing bad pixels due to detector defects and cosmic rays.  It adjusts for seeing variations that occur over the course of an observation, and can also minimize the contamination from nearby stars that happen to fall on the slit.  We have determined that the contamination from the M-dwarf companion $\tau$~Boo\,B is negligible in all epochs of our extracted $\tau$~Boo\,A spectra.  Arc lamps typically used for wavelength calibration do not provide useful reference lines in the thermal infrared.  Instead, we wavelength calibrate our spectra using unblended telluric features with accurately known rest wavelengths.  Between 15 and 30 individual telluric lines are used for each order.

\begin{table}[h]
   \caption{L-band observations of Tau Boo b}
   \begin{tabular}{ccccc}
   \hline
   \hline
Date & JD - 2450000 & Phase (rad) & $V_{bary}$ (km/s) & S/N$_{3\mu m}$ \\
   \hline  
21 May 2011 &5702.8542    & 0.3199 &    -17.36  & 6525  \\
03 Apr 2012 &6021.0625    & 0.3847 &    2.26    & 4768\\
01 Apr 2012 &6019.0625    & 0.7810 &    3.02     & 2871\\
14 Mar 2011 &5634.9375    & 0.8164 &   11.75   & 5467\\
24 Mar 2011 &5644.9375    & 0.8353 &   7.39  & 6575\\
              &              &           &   &         \\
\end{tabular}
\label{table:tau_boo_obs}
 \end{table}
 
 
We first correct the bulk telluric absorption with TERRASPEC \citep{Bender12}, a synthetic forward-modeling algorithm that uses the line-by-line radiative transfer model LBLRTM \citep{Clough05} to generate a synthetic absorption function for the Earth's atmosphere (hereafter, TAF).  The TAF includes continuum absorption and line absorption for 28 molecular species, with line information provided by the HITRAN 2008 database \citep{Rothman09}.  Vertical mixing profiles for the seven most prominent species (H$_2$O, CO$_2$, O$_3$, N$_2$O, CO, CH$_4$, and O$_2$)
can be scaled or adjusted from the LBLRTM default profiles, which include the US Standard 1976, tropical, midlatitude, and subartic models.  These
models represent the average atmosphere for their respective latitudes (in 1976), which is a good initial guess for the instantaneous
atmosphere corresponding to a single observation.  We use the tropical model to provide initial vertical profiles for spectroscopy obtained from Mauna Kea.

TERRASPEC convolves the TAF with an instrumental broadening profile (hereafter, IP), which is measured from the data.  The IP is parameterized as a central Gaussian surrounded by satellite Gaussians offset by a fixed percentage of the Gaussian width, and with adjustable amplitudes.  This is similar to the IP parameterization described by \citet{Valenti95} for use with $\mathrm{I_2}$ absorption cells. Typically 2-4 satellite Gaussians serve to model the IP.  The instrumental broadened TAF is then multiplied by a low-order wavelength dependent polynomial to correct for the combined effects of blaze sensitivity and stellar continuum.  TERRASPEC uses the least-squares fitting algorithm \texttt{MPFIT} \citep{Markwardt09} to optimize the parameters comprising the TAF, IP, and continuum.  The FWHM of the profile is consistently $1.3\times10^{-4} \mu$m, which yields $R = \lambda / \Delta \lambda = 24,000$. Spectral regions containing stellar absorption are excluded from the optimization by adaptive masks.  

\begin{figure}[h]
\begin{center}
\includegraphics[scale=.65]{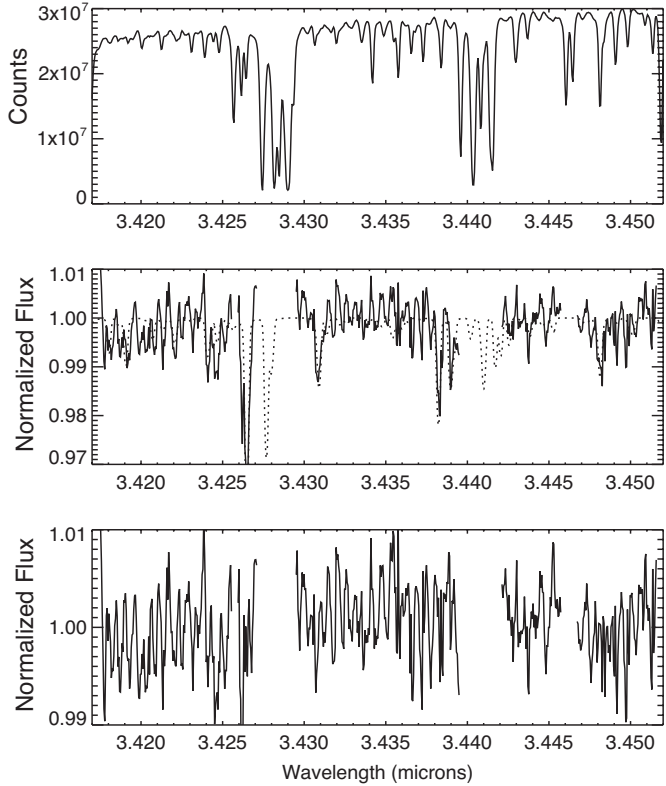}
\end{center}
\caption{The longest wavelength order of data from 14 March 2011. \textit{Top:} The original wavelength-calibrated data.  \textit{Middle:} Telluric-removed spectra with stellar features overlaid (dotted line). \textit{Bottom:} Stellar- and telluric-removed spectrum.}
\label{fig:tell_res}
\end{figure}

The initial telluric correction can be significantly affected by the instrumental fringing.  We therefore mask strong stellar lines from the telluric corrected spectrum, and analyze the remaining spectral range with a Lomb-Scargle periodogram \citep{Scargle82} to measure the frequency and power of the individual fringes present in our spectra.  Two prominent fringes are seen at $\sim$1.75 cm$^{-1}$and 2.18 cm$^{-1}$.  We then calculate the composite fringe function as the product of the individual fringes, and divide it into the non-telluric-corrected spectrum.  The defringed spectrum is then reprocessed with TERRASPEC, yielding the telluric-free and continuum-normalized stellar spectrum, a measurement of the TAF, and a parameterization of the IP.  This process was often iterated 2-3 times to ensure full removal of fringing and telluric absorption. Regions with atmospheric transmission $\leq$40\% of the continuum value, determined from the TAF, are masked and excluded from the final planet search.  Figure \ref{fig:tell_res} demonstrates the telluric removal process.
   
\subsection {Two-Dimensional Cross Correlation}

Next we use a two-dimensional cross correlation analysis, TODCOR \citep{Zucker94}, to simultaneously extract the planetary and stellar velocity shifts.   We generated a synthetic stellar
model of the $\tau$~Boo\,A spectrum using a recent version of the LTE line analysis code MOOG \citep{Sneden73} and the MARCS grid of stellar atmospheres \citep{Gustafsson08}.  The input linelist was created by detailed matching of a synthetic
solar spectrum to the ATMOS ATLAS-3 infrared solar spectrum \citep{Abrams96}, starting from the solar linelist generated by Sauval  \citep[see][]{Hase06}.  Using the MARCS solar
model and solar abundances in \citet{Grevesse07}, adjustments were then made to the atomic line parameters, in particular the
$gf$-values and damping constants, to fit the solar spectrum.  For $\tau$~Boo, a stellar atmosphere with effective temperature T$_{\rm eff}$ = 6375 K, surface gravity $\log{g}$ = 4.0,
and metallicity [m/H] = +0.25 was adopted, based on a review of abundance analyses in the literature. Individual abundances were set by
matching observed lines for elements that were well measured by NIRSPEC (Fe, Si, Mg, Na); otherwise, an abundance
of +0.25 was used.  

A plane-parallel model was calculated for $\tau$~Boo\,b using the PHOENIX stellar and planetary atmosphere code \citep{Barman01, Barman05}.   The planet is very hot, with an equilibrium dayside temperature between 1600--2000~K, depending on the day-to-night redistribution of incident stellar flux.   Only 2$\pi$ redistribution (T$_{\rm eq}$ $\sim$2000\,K) was used here.    With an unknown planet radius, the surface gravity was arbitrarily set to $10^4$ cm$/$sec$^{2}$.  As discussed above, water lines are the primary signal we seek in the planetary spectrum and to accurately model these we use the best available water line list from the ExoMol group \citep{Barber06}.   A final high-resolution spectral template was calculated at 10$\times$ the observed resolution ($\Delta \lambda = 0.05$ {\AA}).  

For every epoch, the target spectrum is cross-correlated to determine the cross-correlation function (CCF) for each order.  The planet/star spectroscopic flux ratio is set to 10$^{-5}$, the same order of magnitude as the expected photometric contrast between the two objects.  We also tested flux ratios from 10$^{-3}$ to 10$^{-7}$ and the shape of the resulting maximum likelihood function remains the same because the analysis is only weakly sensitive to the absolute contrast ratio.

\subsection{Maximum Likelihood Analysis}

To find the most likely solution, each CCF must be converted into a probability, or likelihood, $\mathcal{L}$.  To do this, we start with a relationship between $\mathcal{L}$ and the familiar $\chi^{2}$ statistic
\begin{eqnarray}
\mathcal{L} &=& \frac{1}{\sqrt{2 \pi} \sigma_{i}}\prod_{i}  \exp{ \left(- \frac{ \chi^{2}_{i}}{2}\right)} \nonumber \\
\log \mathcal{L} &=& \kappa - \frac{\chi^{2}}{2}
\end{eqnarray}

\noindent where $\kappa$ is a constant that does not matter when comparing relative likelihoods, assuming $\sigma_{i} =  \sigma = \rm{const}$.
Denoting the observed spectra as $S_i$ and the template spectra as $f_i \equiv f(\lambda_i + \Delta\lambda)$ yields
\begin{eqnarray}
\chi^{2} &=& \sum_i \frac{(S_i - f_i)^2}{\sigma^{2}} \nonumber \\
&=& \frac{1}{\sigma^2} \sum_i (S_i^2 + f_i^2 - 2 S_i f_i)
\end{eqnarray}

\noindent Each part of the quotient can be summed individually. For a continuum-normalized spectrum with N points, this results in
\begin{equation}
\sigma^{2} = \frac{1}{N}\sum S_{i}^2 
\end{equation}
\begin{equation}
\label{eqn:step}
\chi^{2}_{i} = \mathrm{const} - \sum \frac{S_{i} f_i}{\sigma^{2}}
\end{equation}

\noindent The second term on the right-hand-side of Eqn~\ref{eqn:step} is simply the \rm{CCF}.  Thus, the \rm{CCF} and $\mathcal{L}$ are related by:
\begin{equation}
\log \mathcal{L} = \mathrm{const} + CCF
\end{equation}

\begin{figure}[h]
\begin{center}
\includegraphics[scale=.97]{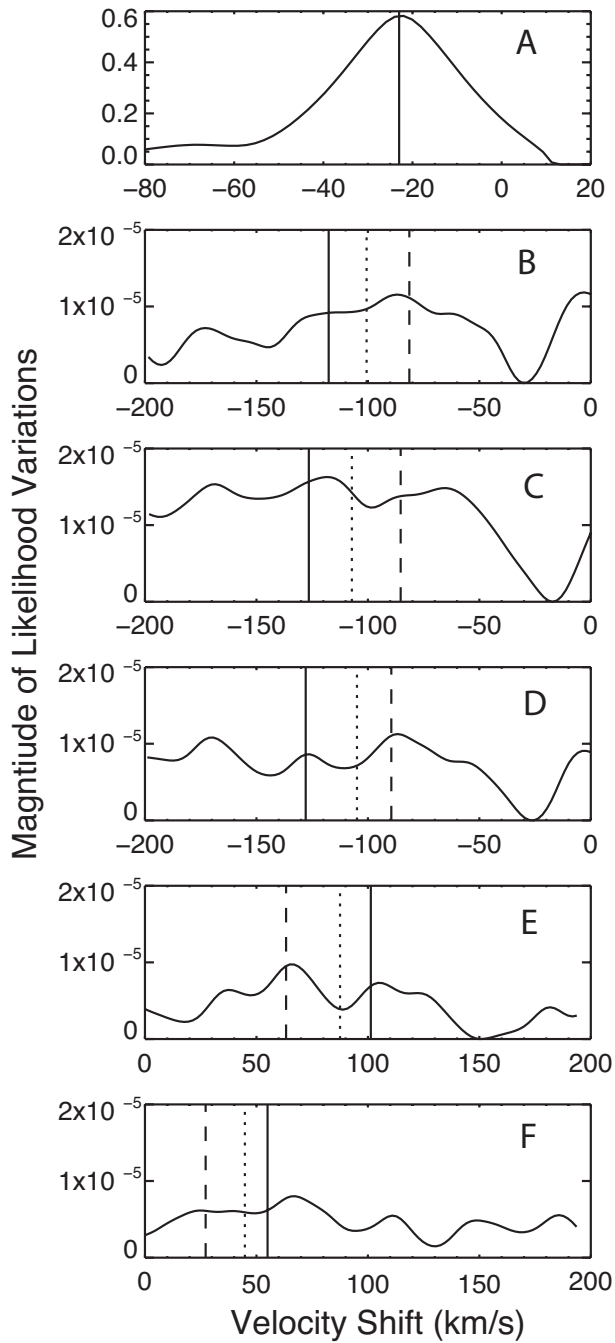}
\end{center}
\caption{\textit{A:} The ML function of the stellar velocity shift using the data from 24 March 2011. \textit{B-F:} The ML function for the planet signal for 24 March 2011, 14 March 2011, 1 April 2012, 21 May 2011, and 3 April 2011, respectively.  Note the changing sign of planetary velocities for each epoch, which are anti-correlated with the sign of the stellar RV shift.  The vertical lines correspond to the velocity shift at a given epoch for an orbital solution with K$_{p}$=70 (dashed), K$_{p}$=90 (dotted), and K$_{p}$=110 km$/$s (solid).}
\label{fig:cc_all}
\end{figure}

The goal is to maximize $\sum \log L$ for both the stellar and planetary velocity shifts, where the sum is over all spectral orders for an individual epoch.  Figure \ref{fig:cc_all} (top panel) demonstrates that the stellar velocity clearly stands out as the most likely solution for one epoch, as is the case for all other epochs as well.  Our NIRPSEC spectra have insufficient RV precision to be sensitive to the orbital motion of $\tau$~Boo\,A so the stellar RV is therefore fixed at the systemic value. Barycentric movement is accounted for.  

The maximum likelihood (ML) solution of the planet's orbit is much more complex. The likelihood is proportional to the CCF, a two-dimensional surface that reflects coherence in features between template spectra and those of the target.  Due to the multiplicity of rovibrational transitions in the asymmetric top spectrum of hot water, the correlation coefficient remains large ($>$0.9) over significant offsets.  This results in multiple ``peaks'' at incorrect velocity lags, as can be seen in Fig. \ref{fig:cc_all}.  Thus, a single epoch would lead to a degeneracy of solutions.  Fortunately, we do not worry about alignment between planet and telluric spectra because not only are the spectra very different, but the combined systemic and planetary RV shifts the planetary spectrum sufficiently to avoid collisions between spectral features. 

\subsection{Orbital Solution}

To break the degeneracy several epochs of data must be brought to bear.  Using the known period of the planet, we calculate the orbital phase of each epoch (Table \ref{table:tau_boo_obs}) and use that to find the most likely planetary velocity consistent with a circular Keplerian orbit (the estimated planetary eccentricity is small, e $\sim$0.02, \citep{Butler97, Brogi12}).
Since the absolute orbital velocity of the planet is known from the period, we are actually interested in K$_{p}$, or semi-amplitude.  A range of K$_{p}$ is tested, each corresponding to a different inclination of the planet, as well as a unique mass.  Furthermore, each K$_{p}$ leads to a different velocity lag at a given phase, such that 
\begin{equation}
\label{eqn:planetvel}
v_{pl} = \mathrm{K}_{p} \sin(\omega~t + \phi) + \gamma
\end{equation}
where $\omega = 2~\pi / P$, $\phi$ is a phase lag that can be set to zero by choosing the proper starting date, and $\gamma$ is the combined stellar barycentric and systemic velocities ($V \sin{i} $$\approx$ 15 km$/$s \citep{Butler06} and $V_{bary}$ listed in Table \ref{table:tau_boo_obs}) .  We then seek the best-fitting K$_p$ by maximizing the sum of the likelihood of $v_{pl}$ for all epochs, whose highest value indicates the most likely solution for K$_{p}$.

\section{Results}

  It is clear that the ML curve for each epoch in Figure \ref{fig:cc_all} shows multiple peaks.  Only one of these peaks per epoch can be the true peak that results from the planet signal in our spectra; the other peaks are artifacts caused by chance misalignments.  The molecular lines in the model water spectrum can align by chance with both signal and noise features in the observed spectra and with other molecules in the planetary atmosphere that have not been included in the planetary template.  However, the true peaks can be distinguished from artifacts by requiring a multi-epoch orbital solution that is consistent with a Keplerian orbit.  The dashed, dotted, and solid lines in Fig. \ref{fig:cc_all} represent the $v_{pl}$ for three Keplerian solutions with K$_{p}$ = 70 km$/$s, 90 km$/$s, and 110 km$/$s, respectively.  While none correspond to the highest peak for all epochs, certain orbital solutions find peaks more often than troughs.  

\begin{figure}
\begin{center}
\includegraphics[scale=.85]{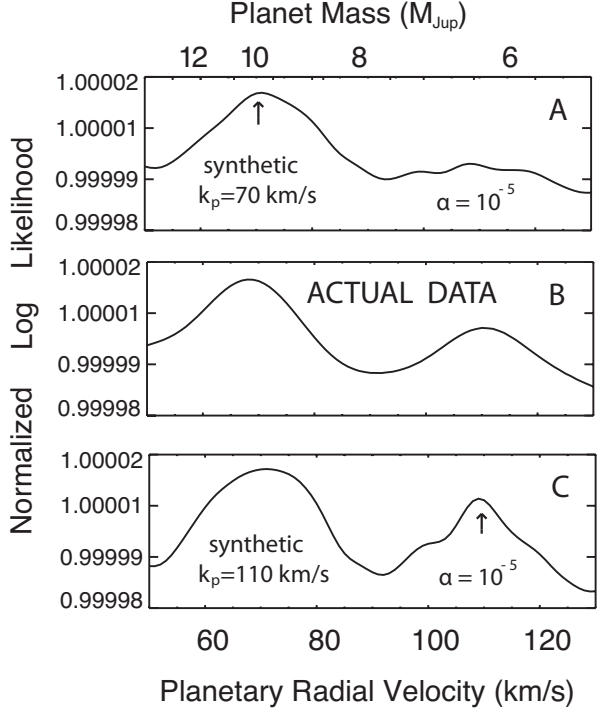}
\end{center}
\caption[Prior]{The normalized log likelihood as a function of planetary velocity, K$_{p}$. \textit{A:} Results from synthetic spectra, composed of the stellar and planetary templates, with a planetary signal injected at 70 km$/$s and analyzed with the same procedures applied to the data. \textit{B:} The data analyzed using a planet-to-star flux ratio of 10$^{-5}$ for a water vapor around a planet with $T_{eq}$ $\sim$2000 K.  \textit{C:} Same as \textit{A} but for a signal injected at 110 km$/$s.}
\label{fig:ML_kp}
\end{figure}

 To find the most likely orbit, the likelihoods at every epoch for a given K$_p$ are combined. The second panel in Figure \ref{fig:ML_kp} presents this composite maximum likelihood, along with the corresponding planetary mass.  This function represents the sum of the log likelihoods of all five epochs.  There are two peaks in the likelihood function, at K$_{p}\sim$ 70 km$/$s and 111 km$/$s.  We have demonstrated empirically the degenerate effects of cross-correlating a water spectrum.  Now we explore the systematics of the analysis and consider how both the properties of the individual stellar and planetary template autocorrelation functions, as well as the correlation of the two template spectra, can affect the composite likelihood function. 

To do this, synthetic data sets are subjected to the same analysis as the original data.  Synthetic target spectra of just the stellar and planetary templates are used, each shifted to the correct velocity for a given epoch.  The results are given in Fig. \ref{fig:ML_kp}, whose top panel shows the normalized log likelihood from a planetary spectrum injected at 70 km$/$s, with a planet-to-star flux ratio of 10$^{-5}$. The proper signal is clearly retrieved.  However, the third panel shows that at an injected planetary radial velocity of 110 km$/$s and the same contrast ratio, a double-peaked log likelihood results.  Thus, following exactly the same procedure as was used for the observations, a ``perfect'' target spectrum with no noise and no (terrestrial) atmosphere retrieves both the correct and an additional signal, with a structure that mimics the data, likely as a result of the complex hot water spectrum near 3 $\mu$m.  

\begin{figure*}
\begin{center}
\includegraphics[scale=.52]{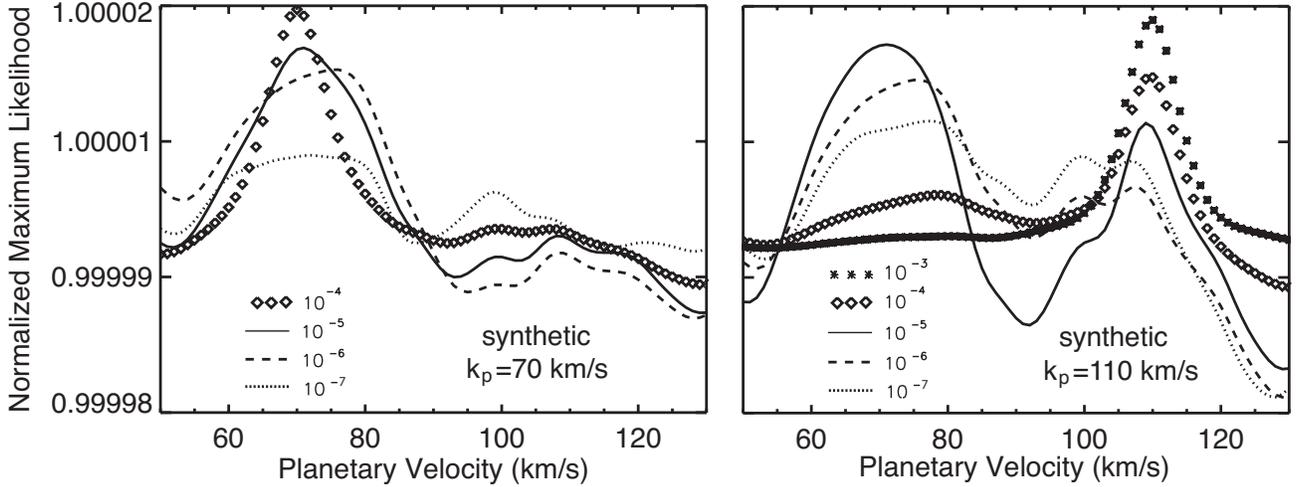}
\end{center}
\caption[Prior]{The normalized log likelihood as a function of planetary velocity, K$_{p}$, for different spectroscopic flux ratios. \textit{Left:} Results from synthetic spectra, composed of the stellar and planetary templates, with a planetary signal injected at 70 km$/$s and analyzed with the same procedures applied to the data. \textit{Right:} Same for a signal injected at 110 km$/$s.  The signal at 110 km$/$s requires a much stronger planetary signal to uniquely determine the correct orbital velocity.}
\label{fig:ML_comp}
\end{figure*}

At a sufficiently high planet-to-star flux ratios, the true planetary signal should dominate the posterior likelihood.  Indeed, using a flux ratio of 10$^{-3}$, \citet{Birkby13} detect water absorption from another (transiting) hot Jupiter, HD\,189733\,b.  The right panel of Fig. \ref{fig:ML_comp} shows that when the flux ratio of the planetary signal is increased to this level, the singular correct velocity is retrieved.  We can constrain the spectroscopic contrast of $\tau$~Boo\,b relative to its host star by comparing the data to these synthetic fits.  The simulations show that for $\alpha \geq$10$^{-4}$, the correct 110 km$/$s solution should present a larger maximum likelihood.  Since our data do not demonstrate this, we can put a 1-$\sigma$ upper limit on the high resolution spectroscopic flux ratio of 10$^{-4}$ at $\sim$ 3.3 microns.  The left panel of Figure \ref{fig:ML_comp} shows that a planetary signal of 70 km$/$s would be uniquely retrieved for all realistic flux ratios.

\section{Conclusions \& Future Work}
 The detection of $\tau$~Boo\,b has been a difficult quest.  Contradictory results have been published over the past 15 years and the difficulty in both thoroughly removing the telluric absorption and also extracting the diminutive planetary signal is not to be underestimated.  Multiple observations at different wavebands improve the validity of the detection, and facilitate a variety of statistical tests to ensure an accurate measurement.  

Here, the validity of the most likely solutions for the RV of $\tau$~Boo\,b are explored.  Of the two prominent velocities that fit the data, one is shown to result from the systematics of the cross-correlation analysis performed on this high flux ratio spectroscopic binary system.  Interestingly, previous velocity studies of the planet in reflected light \citep{CollCam99} and methane \citep{Wiedemann01} retrieved values close to this artifact.  We suggest that residual water vapor in the atmosphere, after the telluric removal, might have been responsible for the false value previously reported from infrared observations. As methods for telluric corrections have improved over the past decade, this problem can be overcome and the correct radial velocity retrieved.
 
Our analysis gives a 6-$\sigma$ detection of the planet at K$_{p}$ = 111 $\pm$ 5 km$/$s for $\tau$~Boo\,b, with a 1-$\sigma$ upper-limit on the 3.3~$\mu$m planet/star spectroscopic flux ratio of 10$^{-4}$.  To determine the significance of our detection, we injected synthetic signals at a variety of planetary velocities and planet-to-star spectroscopic flux ratios and constructed the chi-square surface of the maximum likelihood fits.  The orbital velocity is in good agreement with previous RV amplitude detections via CO by \citet{Brogi12} and \citet{Rodler12}, and our analysis reveals the presence of water vapor in the planet's atmosphere.  Furthermore, using a stellar mass of 1.341$^{+0.054}_{-0.039}$ M$_\odot$ \citep{Takeda07} and a stellar velocity semi-amplitude of 0.4664 $\pm$ 0.0033 km$/$s \citep{Brogi12}, we derive a planetary mass of M$_{P}$ = 5.90~$^{+0.35}_{-0.20}$~M$_{\rm Jup}$ with an orbital inclination of $i$ = 45~$^{+3}_{-4}$~degrees.
 
The technique presented here is in its nascent stages, and the work is by no means complete.  Additional quantitative characterization of the physical properties (temperature, opacity) and composition (and thus estimates of vertical mixing) of $\tau$~Boo\,b's atmosphere will require both significant simulations of the atmospheric radiative transfer and data analysis along with additional data at longer and shorter wavelengths.  Such work is beyond the scope of this letter, but expanded studies of $\tau$~Boo using the same methods described above, but also including molecules such as CH$_{4}$ and CO at longer wavelengths than examined here, are underway.  Indeed, although the mole fractions of methane should be insignificant for $\tau$~Boo\,b, we have searched for the molecule using the methods described herein and have no detection to report in our data.  With careful analysis in the future, the relative abundances of these molecules in the atmospheres of non-transiting exoplanets can be ascertained.   Further applications of this technique, using water and methane as template molecules, to a variety of additional hot Jupiter exoplanets will be reported elsewhere.

\section{Acknowledgments}

The W.M. Keck Observatory is operated as a scientific partnership among the California Institute of Technology, the University of California and NASA, and was made possible by the financial support of the W.M. Keck Foundation.  ACL and GAB gratefully acknowledge support from the NSF GRFP and AAG programs,  JAJ the generous grants from the David and Lucile Packard and Alfred P. Sloan Foundations, and CFB support from the Center for Exoplanets and Habitable Worlds, which is supported by the Pennsylvania State University, the Eberly College of Science, and the Pennsylvania Space Grant Consortium. We thank Jacques Sauval for kindly providing a copy of his solar linelist.  Finally, the authors wish to acknowledge the significant cultural role of the summit of Mauna Kea. 

\end{document}